\documentclass[pra,twocolumn,superscriptaddress]{revtex4}

 \usepackage{epsfig}
 \usepackage{dcolumn}
 \usepackage{bm}
 \usepackage{amsmath}
 \usepackage{amssymb}

\begin{document}
 \title{Gutzwiller approach to the Bose-Hubbard model with random local impurities }

\author{Pierfrancesco Buonsante}
\affiliation{C.N.I.S.M. Unit\`a di Ricerca Politecnico di Torino }
\affiliation{Dipartimento di Fisica, Politecnico di Torino,
             Corso Duca degli Abruzzi 24, I-10129 Torino, Italy}
\author{Francesco Massel}
\affiliation{Dipartimento di Fisica, Politecnico di Torino,
             Corso Duca degli Abruzzi 24, I-10129 Torino, Italy}
\affiliation{Department of Engineering Physics, P.O. Box 5100
02015, Helsinki University of Technology, Finland}
\author{Vittorio Penna}
\affiliation{Dipartimento di Fisica, Politecnico di Torino,
             Corso Duca degli Abruzzi 24, I-10129 Torino, Italy}
\affiliation{C.N.I.S.M. Unit\`a di Ricerca Politecnico di Torino }
\author{Alessandro Vezzani}
\affiliation{Dipartimento di Fisica, Universit\`a  degli Studi di Parma and
             C.N.R.-I.N.F.M., 
             Viale G.P. Usberti 7/a, I-43100 Parma, Italy}

\begin{abstract}
Recently it has been suggested that fermions whose hopping amplitude is
quenched to extremely low values  provide a convenient source of local disorder  for lattice bosonic systems realized in current experiment on ultracold atoms. 
Here we investigate the phase diagram of such systems, which provide the experimental realization of a Bose-Hubbard model whose local potentials are randomly extracted from a binary distribution. Adopting a site-dependent Gutzwiller description of the state of the system, we address one- and two-dimensional lattices and obtain results agreeing with previous findings, as far as the compressibility of the system is concerned. We discuss the expected peaks in the experimental excitation spectrum of the system, related to the incompressible phases, and the superfluid character of the {\it partially compressible phases} characterizing the phase diagram of systems with binary disorder.
In our investigation we make use of several analytical results whose derivation is described in the appendices, and whose validity is not limited to the system under concern.

\end{abstract}

\maketitle

\section{Introduction}
\label{sec:intro}
Since the seminal paper by Fisher {\it et al.} \cite{Fisher_PRB_40_546},
disordered bosonic lattice systems have been the subject of active investigation.
The recent impressive advances in cold atom trapping allowed the experimental 
realization of the prototypal bosonic lattice model, i.e. the
Bose-Hubbard model \cite{Greiner_Nature_415_39}. 
Different techniques have been devised for the introduction of disorder in the system \cite{Fallani_AAMOP_56}, 
such as speckle field patterns \cite{PACK1}, incommensurate bichromatic optical lattices \cite{Fallani_PRL_98_130404}, localized fermionic impurities \cite{Ospelkaus_PRL_96_180403}.
In particular Ref.~\cite{Fallani_PRL_98_130404} provides experimental evidences of  the hallmark phase of the disordered Bose-Hubbard model, i.e. the compressible and non superfluid {\it Bose-glass} \cite{Fisher_PRB_40_546}.

At the theoretical level very diverse techniques have been employed in the study of the disordered Bose-Hubbard model. A non exhaustive list includes field-theoretical techniques \cite{Fisher_PRB_40_546,PACK2} quantum Monte Carlo  simulations \cite{PACK3}, mean-field schemes \cite{Sheshadri_PRL_75_4075,Damski_PRL_91_080403,Krutitsky_NJP_8_187,Buonsante_PRA_76_011602,Buonsante_LP_17_538,Buonsante_JPB_40_F265,Bissbort_arXiv_0804_0007} and others \cite{PACK4,Freericks_PRB_53_2691}.

Here we are interested in the case of fermionic impurities.
Bose-Fermi systems have been studied by several Authors, \cite{PACK5}.
If the kinetic energy of the fermionic atoms is negligible, e.g. due to a strong suppression of the relevant hopping amplitude, the impurities localize at random
sites of the optical lattice \cite{Ospelkaus_PRL_96_180403,Gavish_PRL_88_170406}.  The system can be hence described by a Bose-Hubbard model with random local potential characterized by a binary distribution. 
Several years after an early discussion of this model \cite{Allub_SSC_99_955} the features, new features induced in the phase diagram of the system  by binary disorder has been discussed in recent Refs.~\cite{Fehrmann_OptComm_243_23,Mering_PRA_77_023601,Krutitsky_PRA_77_053609,Refael_PRB_77_144511}. A characteristic feature of such a phase diagram consists in the presence of noninteger-filling incompressible lobes.
Mering and Fleischhauer \cite{Mering_PRA_77_023601} provide simple arguments showing that the  phase diagram of the disordered model does not depend on the impurity density and can be straightforwardly derived by that of the homogeneous case, at least as far as compressibility is concerned. In particular one can recognize {\it fully compressible}, {\it fully incompressible} and {\it partially compressible} regions. While the first and the second are clearly superfluid and insulating, respectively, the question arises about the superfluidity of the {\it partially compressible} regions, at least on high dimensional lattices. Indeed, as discussed in Refs.~\cite{Mering_PRA_77_023601,Krutitsky_PRA_77_053609}, the partially compressible phase is bound to be insulating, and hence {\it Bose-glass}, on 1D lattices.

In this paper we describe the zero-temperature mean-field phase diagram of the Bose-Hubbard model with binary-distributed disorder. First of all we show that the above compressibility scenario independent of the impurity density \cite{Mering_PRA_77_023601} is confirmed also by our site-dependent Gutzwiller approach. Moreover the analytical tractability and the computational affordability of this technique allows us to investigate the superfluidity of the partially compressible phase both in one and two dimensional systems. 
In particular, we address the issue of quantum percolation which, as already pointed out in Ref.~\cite{Ospelkaus_PRL_96_180403}, is expected to play a crucial role in this problem. While we confirm that on one-dimensional systems the partially compressible phase is substantially insulating, we find that in higher dimensions the system always exhibits a finite superfluid fraction due to quantum tunneling. However, this superfluid fraction can be so small that the system can be considered virtually insulating. 
Although phase diagrams make rigorously sense in the thermodynamic limit, it should be taken into account that linear dimension of current experimental realizations of the system under concern is of the order of a few hundred sites. It is hence important to consider finite-size effects, which we demonstrate to be quite relevant especially in the partially compressible phases, and to depend significantly on the impurity density.  

The plan of the paper is as follows. In Sec.~\ref{sec:model} we describe the model under investigation and introduce the superfluid fraction as an important parameter in the characterization of the phase diagram thereby. In Sec~\ref{sec:MF} we recall the site-dependent Gutzwiller approach and provide analytic expression for the superfluid fraction and the flux induced by an infinitesimal velocity field in this framework. Section \ref{sec:res} is devoted to the phase diagram of the system. First of all we discuss how the compressibility scenario argued in Ref.~\cite{Mering_PRA_77_023601} is captured by the site-dependent mean-field approach. In Sec.~\ref{s:Mott} we provide an analytical form for the boundaries of the {\it fully incompressible} insulating lobes. Moreover we discuss how the excitation spectrum of the system \cite{Fallani_PRL_98_130404} is modified by the presence of the noninteger insulating phases characterizing of the Bose-Hubbard model with binary disorer. 
Section \ref{s:parc} discusses the superfluidity of the partially compressible phase, in relation to the quantum percolation phenomenon. Finally finite-size effects are investigated in Sec.~\ref{sec:fse} for 1D and 2D systems.
This paper also contains a rich Appendix section where several interesting analytical results are provided for the site-dependent Gutzwiller approach. In particular in Sec.~\ref{sec:App0} we clarify the connection between the mean-field Hamiltonian and the dynamical Gutzwiller equations. In Sec.~\ref{sec:MottA} the analytic formula for the boundaries of the incompressible lobes is derived explicitly. Also we provide an useful inequality and clarify the connection between such a formula and similar results derived in single site mean-field approaches \cite{Fisher_PRB_40_546,Freericks_PRB_53_2691,Krutitsky_NJP_8_187,Fehrmann_OptComm_243_23,Bissbort_arXiv_0804_0007}.
Finally, the superfluid fraction and the flux across neighbouring sites is derived analytically as a function of the mean-field order parameters alone in Sec.~\ref{sec:AppC}. A particularly simple expression applying in the 1D case is also provided.

\section{The  model}
\label{sec:model}
The system under investigation is described by the Bose-Hubbard
Hamiltonian
\begin{equation}
  \label{eq:BHD}
  H  = \frac{U}{2}\sum_{j=1}^{M}n_j(n_j-1)
   +\sum_j^M v_j n_j 
-  t \sum_{i,j} A_{i,j} a_i^\dagger a_j.
\end{equation}
The on-site bosonic operators $a_j$, $a_j^\dag$ and $n_j = a_j^\dag
a_j$ destroy, create and count particles at lattice site $j$,
respectively.  The geometry of the $M$-site lattice is described by the
adjacency matrix $A$, whose generic element $A_{i\,j}$ equals $1$ if
sites $i$ and $j$ are nearest neighbors, and $0$ otherwise. The
parameters $U$ and $T$ are the on-site repulsive strength and the
hopping amplitude across neighboring sites, and, from the experimental
point of view, they are related to the scattering length of the alkali
atoms forming the bosonic gas and the strength of the optical
lattice. We will be considering a binary random distribution for the
local potential $v_j$, namely
\begin{equation}
\label{eq:vj}
p(v_j) = p_0 \delta(v_j-\Delta) + (1-p_0) \delta(v_j)
\end{equation}
This choice is meant to account for the presence of $N_{\rm imp} = M p_0$ 
atoms of a second species trapped at randomly determined sites
 by a strong quench in the relevant hopping
amplitude \cite{Ospelkaus_PRL_96_180403,Gavish_PRL_88_170406}.  The
parameter $\Delta$ measures the strength of the interaction between
these {\it frozen} impurities and the bosons described by Hamiltonian
\eqref{eq:BHD}. In most of the following discussion we assume $\Delta< U$.
The general case can be worked out straightforwardly, and it is briefly 
discussed in Sec.~\ref{s:Mott}.

As it is well known \cite{Fisher_PRB_40_546}, on a homogeneous lattice,
$v_j=0$, the zero-temperature phase diagram of the BH model described
by Eq. \eqref{eq:BHD} comprises an extended superfluid (SF) region and
a series of Mott-insulator (MI) lobes.  The SF phase is gapless,
compressible and characterized by nonvanishing superfluid fraction.  
Conversely, the MI phase is gapped,
incompressible, and characterized by vanishing superfluid and
condensate fractions.  The presence of random potentials is expected
to induce a further {\it Bose-glass} (BG) phase which, similar to MI
is not superfluid, but, similar to SF, is gapless and compressible
\cite{Fisher_PRB_40_546}.  Recently it has been shown that in the case of uniformly box-distributed disorder such a phase
can be captured by a multiple-site mean-field approach
\cite{Buonsante_PRA_76_011602,Buonsante_LP_17_538,Buonsante_JPB_40_F265},
 both on one-
\cite{Buonsante_PRA_76_011602,Buonsante_LP_17_538} and two-dimensional
lattices \cite{Buonsante_JPB_40_F265}. In the latter case the presence
of the harmonic trapping potential typical of experimental systems was
also taken into account. 


  The superfluid fraction is estimated as the response of
the system to an the infinitesimal velocity field imposed on the
lattice.  In the general case such a field is described by the
antisymmetric matrix $B_{i\,j} = -B_{j\,i}$ having nonzero elements
only across neighboring sites. In the reference frame of the moving
lattice the Hamiltonian of the system has the same form as
Eq. \eqref{eq:BHD} except that $A_{i\,j}$ is substituted by $A_{i\,j}
\exp(i \theta B_{i\,j})$, where $\theta$ is a scalar related to the
modulus of the velocity field \cite[see
e.g.][]{Wu_PRA_69_043609,Bhat_PRL_96_060405}. The superfluid fraction
is often defined as the stiffness of the system under the phase
variation imposed by the velocity field \cite[see
e.g.][]{Shastry_PRL_65_243,Roth_PRA_67_031602}

\begin{equation}
\label{eq:sf1}
f_{\rm s} = \lim_{\theta \to 0} \frac{E_\theta-E_0}{t N\theta^2 }
\end{equation}
where $E_\theta$ and $E_0$ are the ground-state energies of the system when 
the lattice is moving and at rest, respectively, while $N$ is the total
number of bosons in the system.
It should be noticed that Eq. \eqref{eq:sf1} is properly a {\it
  fraction}, i.e. a quantity with values in the interval $[0,1]$, only
in simple situations, such as a homogeneous velocity field. More in
general, the condition $\max(|B_{i\, j}|)\leq 1$ ensures that the
superfluid fraction does not exceed 1. 
The imposition of a velocity field induces a flux
across neighboring sites $i$ and $j$ of the form
\begin{equation}
\label{eq:J1}
{\cal J}_{i\,j} = i t A_{i j}\langle\Psi |e^{-i\theta B_{ij}}a_j^\dag a_i-e^{i\theta
  B_{ij}}a_i^\dag a_j|\Psi \rangle,
\end{equation} 
which clearly vanishes for $\theta=0$. In the following we will show that
$f_{\rm s}=0$ only if ${\cal J}_{i\,j}=0$ for any pair of neighbouring sites.
\section{Mean-field approximation}
\label{sec:MF}
The results we are going to illustrate are obtained in the 
widely used site-decoupling mean-field approximation \cite{Krauth_PRB_45_3137,Sheshadri_EPL_22_257}.
 Despite this approach cannot capture the correct behavior
of the spatial quantum correlations, it provides a qualitatively
satisfactory picture of the phases of strongly correlated systems,
even in the presence of spatial inhomogeneities arising from
the harmonic confinement typical of experiments \cite{PACK6} 
or from superimposed disordered potentials \cite{Sheshadri_PRL_75_4075,Buonsante_PRA_76_011602,Buonsante_JPB_40_F265}.

In the strongly correlated regime the state of the system is expected
to be well approximated by a  {\it Gutzwiller} product state
\begin{equation}
  \label{eq:MFgs}
  | \Psi \rangle = \bigotimes_j | \psi_j \rangle, \qquad | \psi_j \rangle = \sum_{\nu=0}^\infty c_{j\,\nu} \frac{\left(a_j^\dag\right)^\nu}{\sqrt{n!}} |\Omega \rangle, 
\end{equation}
where $|\Omega\rangle$ is the vacuum state, $a_j | \Omega \rangle =
0$.  A time-dependent variational principle similar to that
illustrated in Ref. \cite{Amico_PRL_80_2189} results in a set of
nonlinear dynamical equations for the expansion coefficients
$c_{j\,\nu}$ \cite{Jaksch_PRL_89_040402}.  It can be shown that
finding the minimum-energy stationary state (fixed-point) of such
equations is equivalent to finding the ground state of the {\it
  mean-field} Hamiltonian 
\begin{eqnarray}
{\cal H} & = & \sum {\cal H}_i + t \sum_{i \, j} \alpha_i^* A_{i,j} \alpha_j
e^{i\theta B_{i,j}} \label{E:MFH} \\
{\cal H}_i &=& \frac{U}{2} a_i^\dag a_i^\dag a_i a_i + (v_i -\mu) a_i^\dag a_i \nonumber\\
&-& t (\gamma_i a_i^\dag + \gamma_i^* a_i).\label{E:MFH2}
\end{eqnarray}
subject to the self-consistent condition
\begin{equation}
\label{E:SCc}
\gamma_i =  \sum_{j} A_{i,j} \alpha_j e^{i \theta B_{i,j}}, \;\;
 \alpha_i = \langle \Psi| a_{i} |\Psi\rangle = \langle \psi_{i}| a_{i} |\psi_{i}\rangle
\end{equation}
Such a mean-field Hamiltonian is usually derived by introducing the decoupling assumption $a_i^\dag a_j \approx a_i^\dag \alpha_j + \alpha_i^* a_j - \alpha_i^* \alpha_j $ in Hamiltonian \eqref{eq:BHD} \cite{Sheshadri_EPL_22_257}. These issues will be briefly discussed in Appendix \ref{sec:App0}.  
The parameter $\mu$ appearing in \eqref{E:MFH2} is the so-called {\it
  chemical potential}, which comes about due to the fact that the
mean-field Hamiltonian \eqref{E:MFH} does not preserve the total number of bosons, unlike Eq.~\eqref{eq:BHD}.

The mean-field order parameters $\alpha_i$ (with $\theta=0$) allow the characterization of the quantum phases of the system, as far as compressibility is concerned. In particular, on a homogenous system the  (site-independent) $\alpha_i$ is zero  in  the incompressible insulator and finite in compressible superfluid phase  \cite{Sheshadri_EPL_22_257}. On inhomogenous lattices a further situation can in principle occur, where $\alpha_i \neq 0$ only on a fraction of the lattice sites. Following Ref. \cite{Mering_PRA_77_023601} we will define such situation as {\it partially compressible}, as opposed to the {\it fully compressible} and {\it fully incompressible} phases corresponding to the MI and SF.

We observe that in this framework the one-body density matrix has a rather simple expression,
\begin{equation}
\rho_{i\,j} = \frac{\langle \Psi |a_i^\dag a_j|\Psi \rangle}{N} = \frac{\delta_{i\, j} \langle \psi_j |n_j|\psi_j\rangle + (1-\delta_{i\,j}) \alpha_i^* \alpha_j }{N}
\end{equation}
so that the condensate fraction, i.e. the largest eigenvalue of $\rho_{i\,j}$ \cite{Penrose_PR_104_576}, can be estimated as 
\begin{equation}
\label{eq:fc}
f_{\rm c}\propto \frac{1}{N} \sum_i{|\alpha_i|^2}.
\end{equation}
Such a form shows that, at the mean field level, the condensate fraction vanishes only for {\it fully uncompressible} MI phases. Actually $f_{\rm c}$ has been used a convenient order parameter in Refs \cite{Buonsante_PRA_76_011602,Buonsante_LP_17_538,Buonsante_JPB_40_F265,Louis_JLTP_148_351}.

As one expects, the presence of the velocity field involved in the
evaluation of the superfluid fraction modifies the self-consistently
determined mean-field parameters defined in Eq. \eqref{E:SCc}. 
It is easy to show that in the homogeneous case  the superfluid
fraction defined in Eqs.~\eqref{eq:sf1} equals the
right hand side of Eq. \eqref{eq:fc}. That is, the superfluid and condensate
fraction coincide and can be equivalently employed for characterizing the
phase diagram of the system.
In the general case a perturbative approach, carried out in Appendix \ref{sec:AppC}, shows that the superfluid fraction is
\begin{equation}
\label{eq:sf1mf}
f_s=\frac{1}{2N}\sum_{i,j} A_{ij}\alpha_i^0 \alpha_j^0 (B_{ij}-\phi_i+\phi_j)^2
\end{equation}
where the real and positive  $\alpha_j^0$ are the mean-field order parameters for $\theta=0$, and the real phase factors $\phi_j$ depend on the $\alpha_j^0$ according to Eq.~\eqref{e10}.
At the first perturbative order in $\theta$, the mean-field order parameter are
\begin{equation}
\label{eq:al1}
\alpha_j=\alpha_j^0\exp(i\theta \phi_j),
\end{equation}
while the flux defined in Eq.~\eqref{eq:J1} becomes
\begin{equation}
\label{eq:MFJ}
{\cal J}_{i\,j} = 2\theta t \alpha_i^0 \alpha_j^0 A_{i j}[B_{i\,j}-\phi_i +\phi_j]
\end{equation}
Clearly the superfluid fraction  in Eq.~\eqref{eq:sf1mf} vanishes only if each of the terms in the sum vanishes, i.e. if all of the fluxes in Eq.~\eqref{eq:MFJ} are zero. Expectedly, this happens in the MI phase, where $\alpha_j^0=0$ at every site. The same can happen under more general conditions. Indeed, it is sufficient that
$\phi_i-\phi_j=B_{i,j}$ whenever $\alpha_i^0\alpha_j^0 \neq 0$.  This
is precisely what happens in a BG phase, where $f_{\rm s} = 0 $ despite
the system is  compressible.

It is interesting to note that on 1D systems, owing to  the conservation of 
flux, the evaluations of Eqs. \eqref{eq:MFJ} and \eqref{eq:sf1mf} 
does not require the determination of the phases $\phi_j$ via Eq. \eqref{e10}. Indeed, as illustrated in Appendix \ref{sec:AppC}, one gets
\begin{equation}
\label{eq:J1D}
{\cal J}_{j\, j+1}= -{\cal J}_{j+1\, j} = {\cal J} = 2 t \theta \left(\sum_\ell \frac{1}{\alpha_\ell^0 \alpha_{\ell+1}^0}\right)^{-1}
\end{equation}
and
\begin{equation}
\label{eq:fs1D}
f_s= \frac{\cal J}{2 \theta t N}
\end{equation}
It is clear from Eq.~\eqref{eq:J1D} that ${\cal J}=0$ and $f_{\rm s}=0$ as soon as one of the local mean-field parameters vanishes. This explicitly shows that the partially compressible phase is insulating, and hence {\it Bose-glass}, in 1D system, as already mentioned \cite{Mering_PRA_77_023601,Krutitsky_PRA_77_053609}.
\section{Phase Diagram}
\label{sec:res}
\begin{figure}
  \centering
  \epsfig{file=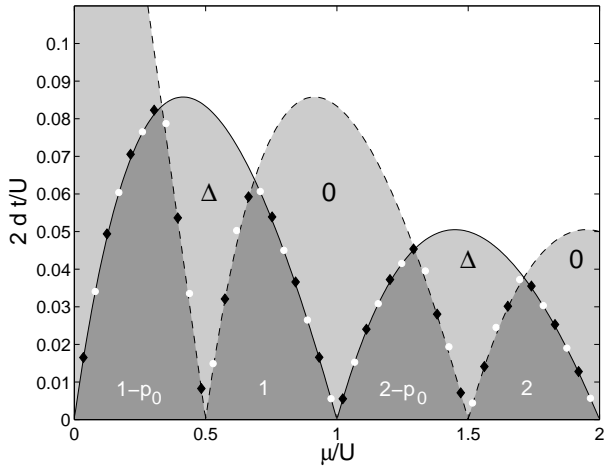,width=9cm}
  \caption{Expected phase diagram for a $d$ dimensional lattice. The solid and dashed black curves are the (mean-field) boundaries involved in Eq.~\eqref{E:mfcb}. These delimit three phases, as far as compressibility is concerned. The {\it uncompressible} lobes (dark gray), the {\it partially compressible} regions (light gray) and the {\it fully compressible} region (white). The uncompressible regions are labeled by the relevant filling. The partially compressible regions are labeled by the local potential of the {\it favourable} sublattice (see text for more details).
The data points have been obtained as described in Sec.~\ref{s:Mott} for a very large 1D lattice ($M= 10^5$) and two impurity densities, $p_0= 0.2$ (white circles) and  $p_0= 0.6$ (black circles). Note that both data sets agree very well with the  expected density-independent analytic result, Eq.~\eqref{E:finc}.}
  \label{fig:Mering}
\end{figure}

As discussed in Ref.~\cite{Mering_PRA_77_023601}, the zero-temperature phase diagram of
the system with binary disorder can be easily inferred from that of 
a homogeneous system, at least as far as compressibility is concerned. Indeed, independent of the impurity density $N_{imp}/M$, in the thermodynamic limit $M\to \infty$ a finite fraction of the disordered system consists of
arbitrarily large regions of uniform local potential (Lifschitz rare regions, see  \cite{Mering_PRA_77_023601,Krutitsky_PRA_77_053609}). The bulk of these regions will behave as a homogeneous lattice, undergoing  a transition at the analytically known critical value \cite{Fisher_PRB_40_546}
\begin{equation}
\label{E:mfcb}
\frac{t}{U} =\frac{1}{2 d}\, {\cal B}\left(\frac{\mu-v}{U}\right), \;\; {\cal B}(x)=\frac{(x-\lfloor x\rfloor)(\lceil x \rceil -x )}{x+1}
\end{equation}
where  $\lfloor x\rfloor$  denotes the largest integer smaller  than $x$,  $\lceil x \rceil=\lfloor x\rfloor+1$, $d$ is the lattice dimension and $v$ is the local potential within the homogeneous regions, which can attain the values $0$ and $\Delta$. 
It is clear that the system is {\it fully compressible} or {\it fully incompressible} only if all of the above region exhibit the relevant property. This means that the region of the phase diagram
\begin{equation}
\label{E:fcom}
\frac{t}{U} > {\cal B}_1\left(\frac{\mu}{U},\frac{\Delta}{U}\right) =  \frac{1}{2 d}\, {\max}\left[
 {\cal B}\left(\frac{\mu-\Delta}{U}\right), {\cal B}\left(\frac{\mu}{U}\right)\right]
\end{equation}
is fully compressible, while the incompressible lobes correspond to the region
\begin{equation}
\label{E:finc}
\frac{t}{U} < {\cal B}_2\left(\frac{\mu}{U},\frac{\Delta}{U}\right)=\frac{1}{2 d}\, {\min}\left[
 {\cal B}\left(\frac{\mu-\Delta}{U}\right), {\cal B}\left(\frac{\mu}{U}\right)\right]
\end{equation}
Fig.~\ref{fig:Mering} shows the phase diagram of the system for  $\Delta=0.5\, U$. The fully compressible and fully incompressible regions correspond to white and dark gray shading. As in the homogeneous case, the fully incompressible lobes correspond to plateaus of the system filling. Interestingly, the binary disorder causes the appearance of non-integer critical fillings of the form $\lceil \mu/U\rceil - p_0$, where $p_0$ is the impurity density \cite{Fehrmann_OptComm_243_23,Mering_PRA_77_023601,Krutitsky_PRA_77_053609}. The incompressible MI phases will be discussed in  detail in Section~\ref{s:Mott}.
The light gray region of Fig.~\ref{fig:Mering}, enclosed between the boundaries ${\cal B}_1$ and ${\cal B}_2$, is of course compressible, yet it contains arbitrarily large incompressible regions. Therefore it is referred to as {\it partially compressible} \cite{Mering_PRA_77_023601}.

This apparently simple scenario requires some clarifications when the possible superfluidity of the system is concerned. As it can be understood from Eq. \eqref{eq:MFJ}, the system can sustain a superfluid flow only along a path where the mean field parameters $\alpha_j$ are not vanishing. Hence, as expected, the system is not superfluid in the incompressible regions of the phase diagram. 
Conversely, in the {\it fully compressible} region the mean-field parameters are nonzero almost everywhere, and a finite superfluid fraction is expected.
The most interesting regions are the {\it partially compressible} ones, which can behave as BG phases, as it will be discussed in Sections~\ref{s:parc} and \ref{sec:fse}.

\subsection{Incompressible phases}
\label{s:Mott}
The boundaries of the incompressible Mott lobes have been derived in Ref.~\cite{Fehrmann_OptComm_243_23} based on a single-site mean-field effective theory. Reference \cite{Mering_PRA_77_023601} reports more quantitative results ensuing from strong-coupling perturbative expansions, which are further supported by density matrix renormalization group simulations. Reference \cite{Krutitsky_PRA_77_053609}  also provides results based on strong-coupling perturbative expansions, as well as on exact diagonalization of small 1D systems.
Here we discuss site-independent mean-field results and show that, unlike effective single-site mean-field theories, they do not depend on the impurity density $p_0$, as it is expected.

We  first of all observe that in  the so-called {\it atomic limit}
$t\to0$ the ground-state of Hamiltonian \eqref{eq:BHD} is a product of
on-site Fock states.  That is, Eq. \eqref{eq:MFgs} applies exactly
with $c_{j\,\nu} = \delta_{j\,\nu_j}$, where $\nu_j =
\max\{0,\lceil\mu-v_j\rceil\}$ and the chemical potential $\mu$ is
determined by the constraint on the total number, $N=\sum_j \nu_j$.
Recalling that the local potential $v_j$ is $\Delta$ at $N_{\rm imp} = p_0\, M$
randomly placed lattice sites and $0$ at the remaining $M-N_{\rm imp}$
sites, it is easy to conclude that the total number of bosons is zero
for $-\infty \leq \mu \leq 0$, and subsequently grows stepwise with
increasing chemical potential.  The staircase function is easily
determined if $0\leq \Delta \leq U$.  In this case the rises of the
steps occur at $\mu_k(x) =  k + x$, with $k=0,1, \ldots, \infty$
and $x = 0,\Delta$.  The height of the steps, i.e. the total
population, is $N= M (k+1) - N_{\rm imp}$ for $\mu_k(0) \leq \mu \leq
\mu_k(\Delta)$ $N= M (k+1)$ for $\mu_k(\Delta) \leq \mu \leq
\mu_{k+1}(0)$. In the first case the wavefunction \eqref{eq:MFgs} is
such that $|\psi_j\rangle = |k\rangle$ at the $N_{\rm imp}$ sites with
$v_j = \Delta$, and $|\psi_j\rangle = |k+1\rangle$ at the remaining
sites, where $|k\rangle = (a_j^\dag)^k |\Omega\rangle/\sqrt{k!} $ is
the local $k$-th Fock state. In the second case $|\psi_j\rangle =
|k+1\rangle$ at every lattice site.

It is straightforward to check that these states diagonalize the
mean-field Hamiltonian \eqref{E:MFH} subject to the self-consistency
constraint \eqref{E:SCc} also for any $t>0$, although they do not
always represent the mean-field ground-state of the system.  As it is
illustrated in Sec. \ref{sec:MottA}, this is true only within the
 regions of the $\mu/U$-$t/U$ phase plane described by $0
\leq t/U \leq |\lambda_{\max} |^{-1} $, where $\lambda_{\max}$ is the
maximal eigenvalue of the matrix 
\begin{equation}
 \label{eq:Ldef}
\Lambda = D\,A, \qquad
 D_{m,m'}=\delta_{m,m'}{\cal B}^{-1}\left(\frac{\mu-v_m}{U}\right),  
\end{equation}
 $A$ is the adjacency matrix of the lattice and the function $\cal B$ appearing in the diagonal matrix $D$ is defined in Eq.~\eqref{E:mfcb}.
The numerical diagonalization of $\Lambda$ at different values of $\mu$ shows that the above discussed plateaus extend over lobe-like regions with alternatively noninteger and integer fillings.  

Since in both cases $\alpha_j=0$ we classify these phases as incompressible Mott insulators. As we have discussed above, the superfluid fraction expectedly vanishes.
 However,
integer- and fractional-filling insulating phases are distinguished by
the correlation with the underlying local potential.  The former are
homogeneous despite the presence of such potential.  The latter are
clearly characterized by a disorder directly related to that in the
location of the impurities described by the local potential $v_j$.
We observe that these two uncompressible phase are expected to exhibit
different excitation spectra, which are relevant experimental quantities
\cite{Fallani_PRL_98_130404}. More to the point, the excitation spectrum of the integer-filling Mott phases will be characterized by three peaks at $U-\Delta$, $U$ and $U+\Delta$. As to the noninteger-filling lobes, the peaks are expected at $\Delta$, $U$ and, for fillings larger than 1, at $2U -\Delta$.

The data points in Fig.~\ref{fig:Mering} have been obtained by evaluating
the maximal eigenvalue of the matrix $\Lambda$ as a function of the chemical
potential for a 1D lattice comprising $M=10^5$ sites. Black and white circles
correspond to different impurity densities, $N_{\rm imp} = 0.2 M$ and $N_{\rm imp} = 0.6 M$. Both data sets show a very good agreement with the analytic result in Eq.~\eqref{E:finc} and, as expected \cite{Mering_PRA_77_023601,Krutitsky_PRA_77_053609}, exhibit no dependence on the impurity density $p_0 = N_{\rm imp }/M$. As briefly recalled in Appendix \ref{sec:MottA}, the same result can be equivalently obtained by studying the matrix $\Gamma = \sqrt{D} A \sqrt{D}$ which, unlike $\Lambda$, is symmetric, $\Gamma=\Gamma^t$.

We emphasize that mean-field results based on effective single-site results can be recovered by averaging the above matrices over disorder. We first of all observe that the disorder-averaged version of $\Lambda$ results in the critical boundaries
\begin{equation}
\label{E:smmf}
\frac{t}{U} = \frac{1}{2 d} \left[p_0 {\cal B}^{-1}\left(\frac{\mu-\Delta}{U}\right) + (1-p_0) {\cal B}^{-1}\left(\frac{\mu}{U}\right)\right]^{-1}
\end{equation}
i.e. precisely the same result obtained by the so-called {\it simple man's} mean-field theory of Ref.~\cite{Fehrmann_OptComm_243_23}. Note that Eq.~\eqref{E:smmf} is indeed a simplified result, in that it depends on the impurity density $p_0$, contrary to expectations \cite{Mering_PRA_77_023601,Krutitsky_PRA_77_053609}. We also observe that equivalent simplified approaches have been adopted in earlier papers \cite{Freericks_PRB_53_2691,Krutitsky_NJP_8_187}, albeit with a different disorder distribution, and date back to the seminal work by Fisher et al. \cite{Fisher_PRB_40_546}. A perhaps more structured effective single-site theory is obtained from the disorder averaged version of the matrix $\Gamma$ matrix defined above. Indeed the boundary of the $\alpha_j=0$ phase ensuing from such matrix in the case of uniformly distributed disorder, $p(v_j) = \Theta(v_j+\Delta/2) \Theta(\Delta/2-v_j)$, is very similar to that provided by the {\it stochastic mean-field theory} described in Ref.~\cite{Bissbort_arXiv_0804_0007}. However, it is quite clear that in the case of the binary distribution in Eq.~\eqref{eq:vj} such boundary again depends on the impurity density $p_0$, just like in Eq.~\eqref{E:smmf}.

In the above general discussion we assume $0< \Delta < U$.
For $\Delta > U$ the arrangement of the first few lobes
changes straightforwardly.  For instance, for $U < \Delta < 2 U$ the
unitary-filling lobe disappears, the basis of the lobe at filling
$1-p_0$ extends in the whole interval $0 \leq \mu \leq U$ and
the interval $U \leq \mu \leq \Delta$ provides the basis for a lobe
with filling $2-2 p_0$. Note that if $p_0 = 1/2$
this last lobe has unitary filling, although this results from
averaging sites at filling $0$ and $2$. Hence the unitary-filling
incompressible phase changes from homogeneous to disordered as
$\Delta$ becomes larger than $U$. This was observed in an early work,
where however the disordered, incompressible insulating phase is
identified as a Bose-glass phase \cite{Allub_SSC_99_955}.

Clearly, the phase diagram of the homogeneous lattice is recovered for
$\Delta = 0$, while the case $\Delta < 0 $ can be mapped on the
repulsive case with the suitable number of impurities, $\Delta \to
|\Delta|$, $N_{\rm imp} \to M - N_{\rm imp}$.

\subsection{Partially compressible phase, Bose-glass}
\label{s:parc}
In order to discuss the situation in the partially compressible regions it proves convenient to introduce the notions of {\it favourable} and {\it unfavourable} sublattices for bosons added to the system, as determined by the competition by the {\it local} boson-boson interaction strength $U$ and the {\it local} impurity potential $\Delta < U$.
Clearly, in the absence of bosons the impurity-free sublattice, $v_j=0$, is energetically favourable. If some bosons are introduced in the system, they will prefer the impurity-free sites, as far as the local energy is concerned.
 However, at some fillings, the presence of bosons in the impurity-free sublattice could make the impurity sites more convenient energetically.
Thus the favourable sublattice coincides with the impurity-free sublattice or with the impurity sublattice depending on the filling or, equivalently, on the chemical potential. Typically,  the impurity  sites are preferred in the partially uncompressible regions to the left  of the integer filling uncompressible lobes (marked by $\Delta$ in Fig.~\ref{fig:Mering}), whereas the impurity-free sites are preferred in the remaining regions (marked by $0$ in Fig.~\ref{fig:Mering}).
We call {\it unfavourable} the sites of the lattice not belonging to the {\it favourable} sublattice. 

The partial compressibility of the regions we are considering arises from the presence of arbitrarily large {\it unfavourable} regions behaving like homogeneous systems. Hence there are arbitrarily large sublattices hosting an uncompressible phase characterized by $\alpha_j=0$. However, owing to the hopping term, not all of the unfavourable sites are characterized by vanishing order parameter. This allows for the formation of a cluster of $\alpha_j>0$ sites spanning the entire lattice, and therefore capable of supporting a superfluid flow, even if the {\it favourable} sites do not percolate throughout the lattice.
\begin{figure}
  \centering
  \epsfig{file=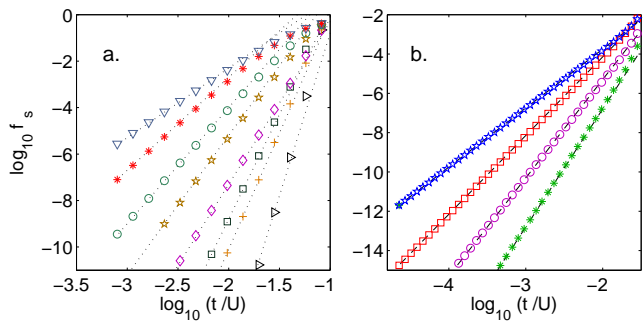,width=0.48\textwidth}
  \caption{Power law decay of the superfluid fraction as a function of $t/U$. The left panel refers to a 1D lattice comprising $M=400$ sites. The different data sets have been obtained by varying the impurity density between $0.4$ and $0.7$. The right panel corresponds to a 2D system, also comprising $M=400$ sites, where the impurity density varies between $0.005$ and $0.1$. In every case the data sets are well described by a straight line having an integer slope which is found to coincide with the {\it percolation length} $L$ (see Fig.~\ref{fig:PL}). In the 1D case (left) we observe lengths from 3 to 14. In the 2D case $L=3, 4, 5, 6$.}
  \label{fig:Qperc}
\end{figure}
That is to say, the superfluidity is not related to the percolation of the favourable sublattice, but rather to the percolation of the sites where $\alpha_j>0$. The latter is made possible by the quantum tunneling effect, which allows for the bridging of possibly disconnected clusters of favourable sites.
\begin{figure}
  \centering
  \epsfig{file=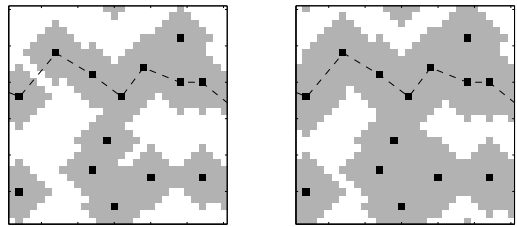,height=3cm}
  \caption{A sketch illustrating the notion of {\it percolation length}. A 30 by 30 square lattice contains 14 favourable sites, signalled by the black squares. Due to their very low density they do not form a cluster spanning the lattice. The dashed line signals the shortest path allowing to ``wade'' through the lattice by stepping on favourable sites. The longest jump has to be taken between the first and the second favourable site from the left. The gray shading demonstrates the effective increase in the density of  favourable sites caused by the long-range connectivity introduced by quantum tunnelling. The range of such connectivity is  4 (left panel) or 5 (right panel) lattice constants. The shading in the rightmost panel shows that the longest jump in the dashed path measures 10 lattice constants. Hence $L=10$ for this favourable sublattice.}
  \label{fig:PL}
\end{figure}
We have studied this phenomenon at $\mu=\Delta=0.5$, i.e. where the partially uncompressible regions of the phase diagram extend down to vanishing hopping amplitude. As it is shown in Fig.~\ref{fig:Qperc}, both in 1D and 2D we observe a behaviour of the form $f_{\rm s} \propto (t/U)^L$, where $L$ is the (integer) {\it percolation length}, i.e. the length of the longest ``bridge'' among those necessary to turn a disjoint impurity distribution into the shortest cluster spanning the whole lattice (see Fig.~\ref{fig:PL}). Note that the same $p_0$ can give a different $L$ due to finite-size fluctuations. This is especially clear on 1D system, where one expects $L=\infty$ in the thermodynamic limit, independent of $p_0$. However, once an $L$ is determined by the actual finite-size realization of the disordered system, it dictates the behaviour of $f_{\rm s}$ as discussed above and demonstrated in Fig.~\ref{fig:Qperc}.

Now on one-dimensional lattices the percolation length corresponds to the size of the largest homogeneous cluster in the unfavourable lattice, which becomes arbitrarily large in the thermodynamic limit, independent of the impurity density. Hence in 1D the partially compressible phase is expected to be insulating for any  finite impurity density as observed in Refs. \cite{Mering_PRA_77_023601,Krutitsky_PRA_77_053609}.

Conversely, on higher dimensional lattices the superfluid fraction is finite at any impurity density, although it can become extremely small. 
This is due to the fact that, unlike the one-dimensional case, $L$ never diverges in the thermodynamic limit. This can be understood by observing that quantum tunneling introduces a long-range connectivity between the possibly disjoint clusters forming the favourable sublattice, which effectively increases the density of favourable sites. It is then clear that a sufficiently large hopping amplitude can bring the effective density of favourable sites above the (finite) percolation threshold, so that an effective spanning cluster is formed  \cite{PACK7}. This concept is illustrated by the light gray regions in Fig.~\ref{fig:PL}.

The above discussion is valid in the thermodynamic limit. As we illustrate in the following, strong deviations from the expected behavior can be observed on finite-size lattices, even for fairly large sizes.
In particular, on 1D systems a finite superfluid fraction can be observed in
the  partially compressible regions of the phase diagram. 
Conversely, on higher dimensional system, the expectedly finite superfluid fraction in these region may become so small that the phase can be considered insulating for any practical purpose. This agrees with the intuitive notion of a {\it glass} as an extremely viscous fluid.

\subsection{Finite-size effects}
\label{sec:fse}
Of course a phase diagram is rigorously defined only in the thermodynamic limit, $M\to\infty$. However, the experimental realizations of the Bose-Hubbard model, based on ultracold atoms trapped in optical lattices, are far from such a limit, especially in the presence of disorder. Indeed, the occurrence of arbitrarily large regions with uniform local potential becomes extremely improbable, if not impossible. The higher the lattice dimension $d$, the more serious this problem. 
Moreover, as demonstrated by Fig.~\ref{fig:Qperc}, the superfluid fraction can be so small that the system can be considered virtually non superfluid. 

\begin{figure}
  \centering
  \epsfig{file=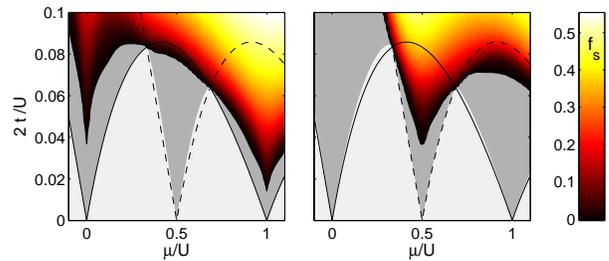,width=8.5cm,bbllx=27,bblly=218,bburx=560, bbury=450,clip=}
\caption{\label{fig:1D_fsz} Phase diagram as determined from a numerical simulation on a 1D lattice comprising $M=961$ sites and containing $N_{\rm imp} = 200$ (left) and $N_{\rm imp} = 600$ (right) randomly located impurities. In both cases $\Delta = 0.5 \,U$, as in Fig.~\ref{fig:Mering}. The density plot represents the superfluid fraction as specified by the colorbar. The dark grey areas enclose the region where $f_{\rm s}$ is smaller than 1\% of its largest value in the entire examined area. The light grey areas are the fully incompressible Mott lobes determined as described in Sec.~\ref{s:Mott}. The solid and dashed black curves are the boundaries involved in Eq.~\ref{E:mfcb}. They are the same as in Fig.~\ref{fig:Mering}.  }
\end{figure}

\begin{figure}
  \centering
  \epsfig{file=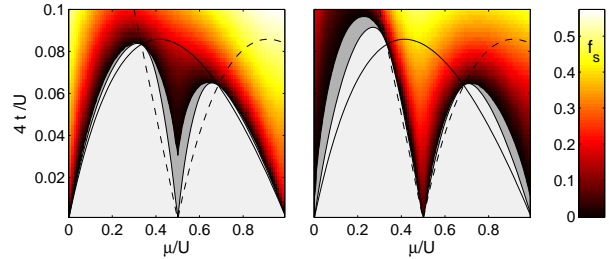,width=8.5cm, bbllx=27,bblly=218,bburx=560, bbury=450,clip=}
\caption{\label{fig:2D_fsz} Phase diagram as determined from a numerical simulation on a 2D lattice comprising $M=31\times 31 = 961$ sites and containing $N_{\rm imp} = 200$ (left) and $N_{\rm imp} = 600$ (right) randomly located impurities. In both cases $\Delta = 0.5 \,U$, as in Figs.~\ref{fig:Mering}.and \ref{fig:1D_fsz}.  }
\end{figure}

In order to demonstrate the relevance of these  effects we have carried out numerical simulations for 1D and 2D lattices comprising $M=961$ sites. In both cases we have adopted periodic boundary conditions, and a constant velocity field parallel to a coordinate direction. The resulting phase diagrams, where we took into account that vanishingly small superfluid fractions can be considered zero, are shown in Figs.~\ref{fig:1D_fsz}  and \ref{fig:2D_fsz}.
The first thing to be noticed is that finite-size effects do not dramatically affect the boundaries of the fully incompressible Mott lobes, especially in one dimension. A slight dependence on the impurity density $p_0$ is observed, at variance with the thermodynamic limit result. In general, as it is explained in Appendix \ref{sec:MottA}, the finite size lobes ``enclose'' those in the thermodynamic limit.
Significant finite size effects are instead evident in the partially compressible phase. As we mention above, in the thermodynamic limit one expects this phase to be insulating in 1D, and superfluid for $d>1$. Actually, as it
is clear from Figs.~\ref{fig:1D_fsz}  and \ref{fig:2D_fsz}, these are  the predominant characters of the partially compressible phases also on finite-size lattices. However, on 1D lattices significant portions of the partially compressible phase are superfluid. Conversely, on 2D lattices small partially compressible regions surrounding the Mott lobes exhibit exponentially small superfluid fractions, so that they can be considered virtually insulating.
While the expected behaviour in the thermodynamic limit is expected to be independent of the impurity density $p_0$, this parameter strongly affects the finite size deviations. In particular, for small impurity densities the partially compressible regions marked $\Delta$ in Fig.~\ref{fig:Mering} tend to be more insulating than thosed marked  $0$. The converse occurs at large impurity densities.

We conclude by emphasizing that the favourable sublattice does not percolate at neither  of the chosen impurity densities. Hence, the fact that $f_{\rm s}>0$ in the partially compressible phases is to be attributed to quantum tunneling effects.

\section{Summary}
\label{sec:conc}

In this paper we address the phase diagram of the Bose-Hubbard model
describing ultracold bosonic atoms loaded in an optical lattice containing
static random local impurities. These are fermionic atoms whose hopping amplitude
has been quenched to extremely low values.
We employ a site dependent Gutzwiller scheme, analyzing both 1D and 2D lattices.
On the one hand we show that this approach confirms that the phase diagram of the
system does not depend on the density of impurities and can be easily derived
from the phase boundary of the homogeneous case, at least with respect to the compressibility of the system. We show that the boundaries of the insulating {\it fully incompressible} region of the phase diagram  are strictly related to the spectral radius of two (block) tridiagonal matrices. We discuss the expected modifications in the  structure of the experimental excitation spectrum of the system occurring due to the presence of the noninteger-filling incompressible lobes appearing in the phase diagram in the presence of binary disorder.
Also we provide exact formulas for the superfluid fraction and the fluxes induced by an infinitesimal velocity field, showing that these quantities ultimately depend on the mean-field order parameters alone. These formulas allow us to investigate the superfluid nature of the {\it partially compressible} regions appearing in the phase diagram of the system owing to the binary disorder.
We discuss the experimentally relevant finite size effects, showing that they strongly depend on the impurity density and mainly affect the boundaries of the partially compressible regions. We show that on one-dimensional system quantum percolation causes the appearance of superfluid domains within these regions. On the other hand, on higher dimensional lattice the in-principle superfluid partially compressible regions contain domains that can be considered virtually insulating, due to the extreme smallness of the superfluid fraction.

Last but not least, in the appendices we provide the explicit derivation of the analytic results employed in the paper, whose validity is not limited to the case of binary distributed disorder. In particular, we discuss the relation the site-dependent Gutzwiller approach and effective single site mean-field theories.
\acknowledgments
The authors wish to thank A. Mering for his extremely valuable comments and suggestions.

The work of P.B. has been supported by the C.N.I.S.M. project {\it Quantum Phase Transitions, Nonlocal Quantum Correlations and Nonlinear Dynamics is Ultracold Lattice Boson Systems}.
The work of F.M. has been partially supported by the M.I.U.R. project {\it Cooperative Phenomena in Coherent Systems of Condensed Matter and their Realization in Atomic Chip Devices}. 

\appendix
\section{Normal modes of the Gutzwiller dynamics}
\label{sec:App0}
As we mention in Sec. \ref{sec:MF}, a variational principle analogous
to that described in Ref.  \cite{Amico_PRL_80_2189} results in the set
of dynamical equations for the expansion coefficients of the
Gutzwiller state \eqref{eq:MFgs} \cite{Jaksch_PRL_89_040402}:
\begin{eqnarray}
\label{eq:Gdyn1}
i \dot c_{j\,\nu} &=& \frac{U}{2} \nu (\nu-1) c_{j\,\nu} + v_j \nu c_{j\,\nu} \nonumber\\
&-& t \left(\gamma_j^* \sqrt{\nu+1}\, c_{j \,\nu+1} + \gamma_j \sqrt{\nu}\, c_{j \,\nu-1}  \right) \\
\label{eq:Gdyn2}
\gamma_j &=& \sum_h A_{j\,h} \alpha_h,\;\, 
\alpha_h = \sum_{\nu=0}^\infty \sqrt{\nu+1}\, c_{j \,\nu}^* c_{j \,\nu+1}
\end{eqnarray} 
It is straightforward to check that the norm of each on-site
Gutzwiller factor, and the (average) total number of bosons in the
system
\begin{equation}
\label{eq:NA}
\langle\psi_j| \psi_j\rangle = \sum_{\nu=0}^\infty |c_{j\,\nu}|^2,\quad N = \sum_j \sum_{\nu=0}^\infty \,\nu\,|c_{j\,\nu}|^2,
\end{equation}
are conserved by the dynamics.

The fixed-point condition $\dot c_{j\, \nu} =0$ for
Eq. \eqref{eq:Gdyn1} results in the set of equations
\begin{eqnarray}
\label{eq:Gfp}
 0 &=& \left[- \epsilon_j + \frac{U}{2} \nu (\nu-1) + (v_j-\mu)\right] \nu c_{j\,\nu} \nonumber\\
&-& t \left(\gamma_j^* \sqrt{\nu+1}\, c_{j \,\nu+1} + \gamma_j \sqrt{\nu}\, c_{j \,\nu-1}  \right)
\end{eqnarray}
where $\mu$ and $\{\epsilon_j\}$ are $M+1$ Lagrange multipliers
ensuring that the total number of bosons and the norms of the
Gutzwiller factors $|\psi_j\rangle$ equal the desired values.

But Eq. \eqref{eq:Gfp} is nothing but the eigenvalue equation for the
on-site mean-field Hamiltonian \eqref{E:MFH2} projected onto the
generic Fock state at site $j$, $|\nu\rangle= (a_j^\dag)^\nu
|\Omega\rangle/\sqrt{\nu!}$,
\begin{equation}
\langle \nu| {\cal H}_j - \epsilon_j |\psi_j\rangle =0
\end{equation}
where we recall that $a_h |\Omega\rangle =0$ for all $h$'s.  Note
indeed that Eq. \eqref{eq:Gdyn2}, which must be considered part of
Eq. \ref{eq:Gdyn1}, is exactly equivalent to the self-consistency
constraint specified by Eq. \eqref{E:SCc}.  By comparing
Eqs. \eqref{eq:Gfp} and \eqref{eq:Gdyn1} we see that the solutions of
the mean-field equations are normal modes of the Gutzwiller dynamics
such that $c_{j\,\nu}(t) = e^{-i t (\epsilon_j +\mu\, \nu)}
c_{j\,\nu}(0)$.  The fixed-point nature of these solutions becomes
clear when one considers the relevant expectation values on Hermitian
operators, corresponding to observable quantities. It is easy to
verify that number-conserving Hermitian operators, such as for
instance $a_k^\dag a_h + a_h^\dag a_k $, produce time-independent
expectation values, whereas non number-conserving Hermitian operators,
such as $a_h^\dag + a_h$, produce expectation values oscillating
around 0.  Note that the latter should be identically zero, since the
original Hamiltonian \eqref{eq:BHD} --- unlike its mean-field
counterpart --- commutes with the total number of bosons. This result
is recovered after time-averaging the expectation values.

Note finally that $E= \langle\Psi| H| \Psi\rangle = \langle\Psi| {\cal
  H}+\mu N|\Psi\rangle$, and that ${\cal H}$ can be obtained from
$H-\mu N$ by assuming that $a_j^\dag a_k = a_j^\dag \alpha_k
+\alpha_j^* a_k -\alpha_j^* \alpha_k $ for $j\neq k$, where $\alpha_j
= \langle\Psi| a_j| \Psi\rangle = \langle\psi_j| a_j| \psi_j\rangle$
\cite{Sheshadri_EPL_22_257}.

As it is shown in Refs.~\cite{Amico_PRL_80_2189,PACK8} the TDVP approach based on Glauber's and SU($M$) coherent states instead of those in \eqref{eq:MFgs} gives the discretized Gross-Pitaevskii equations for the Bose-Hubbard model \eqref{eq:BHD}. Interestingly, the Gutzwiller mean-field states in Eq.  in \eqref{eq:MFgs} reduce to Glauber's coherent states for $U\to0$, as it is clear from the form of the mean-field Hamiltonian \eqref{E:MFH} \cite{Jaksch_PRL_89_040402}.


\section{Mott Phase Boundary}
\label{sec:MottA}
In this appendix we show that the critical boundary of the (mean-field) 
Mott phase, ($\alpha_j=0$ for every $j$) is the inverse of the maximal eigenvalue of the 
matrix $\Lambda$ defined in Eq.~\eqref{eq:Ldef}.

Since $\alpha_j=0$ everywhere, we also have $\gamma_j=0$ at every site, according to Eq.~\eqref{E:SCc}. Hence the mean-field Hamiltonian \eqref{E:MFH} is the sum of the on-site Hamiltonians in Eq.~\eqref{E:MFH2} and the 
the ground state \eqref{eq:MFgs} is bound to be a product of local Fock states
\begin{equation}
\label{E:gsAc2}
  |\psi \rangle = |\nu\rangle = \frac{(a^\dag)^\nu}{\nu!}|0\rangle,
  \quad \nu = \lceil \frac{\mu - v}{U} \rceil,
\end{equation}
The relevant on-site energy is
\begin{equation}
\label{E:lgsE}
\epsilon_\nu = \frac{U}{2} \nu(\nu-1) + (v-\mu) \nu  
\end{equation}
Hence $\langle \Psi|a_j|\Psi \rangle = 0$ at every site, and the
self-consistency constraint (\ref{E:SCc}) is satisfied. 
Note that Eq.~\eqref{E:SCc} defines a map, since any set of 
(possibly nonzero) $\alpha_j$ determine a set of local ground states $|\psi_j\rangle$ via Hamiltonian~\eqref{E:MFH}, which in turn determine a new set of $\alpha_j$. This is by definition a fixed point of the map when it coincides with the original set, which is exactly what happens for the configuration under examination, $\alpha_j=0$, for any choice of the Hamiltonian parameters.
However, the stability of such ``trivial fixed point'' does depend on the 
Hamiltonian parameters. Specifically, the fixed point is stable only
if the maximal eigenvalue of the linarized version of the map is smaller than 1.
In order to linearize the map we assume that $|\alpha_j| \ll 1$, which
yields $|\gamma_j| \ll 1$ and treat the
(mean-field) kinetic term in Hamiltonian (\ref{E:MFH}) as
perturbative.  Dropping for a while the site label we get, up to the first  perturbative order,
\begin{equation}
|\psi\rangle = |\psi^{(0)}\rangle + |\psi^{(1)}\rangle
\end{equation}
\begin{eqnarray}
|\psi^{(1)}\rangle = -t \sum_{\nu' \neq \nu}
 \frac{\langle \nu'|\gamma a^\dag + \gamma^* a|\nu \rangle}
      {\epsilon_\nu -\epsilon_{\nu'}} |\nu'\rangle
\end{eqnarray}
\begin{equation}
|\psi\rangle = |\nu\rangle -  \frac{t \gamma \sqrt{\nu+1}}
                                 {\epsilon_\nu -\epsilon_{\nu+1}} |\nu+1\rangle -
                            \frac{t \gamma^* \sqrt{\nu}}
                                 {\epsilon_\nu -\epsilon_{\nu-1}} |\nu-1\rangle  
\end{equation}
where $|\nu \rangle$, $\nu$ and $\epsilon_\nu$ are defined in
Eqs.~(\ref{E:gsAc2}) and (\ref{E:lgsE}).  Hence
\begin{eqnarray}
\langle \psi| a |\psi \rangle  &=&  -t \gamma \left(\frac{\nu+1}{\epsilon_\nu -\epsilon_{\nu+1}} +\frac{\nu}{\epsilon_\nu -\epsilon_{\nu-1}}\right) \nonumber\\
& = & - \gamma \frac{t(U+\mu-v)}{(\nu U+v-\mu)(\nu U-U+v-\mu)}
\end{eqnarray}
Restoring the site label, recalling the definition of $\gamma_m$,
Eq.~(\ref{E:SCc}), $\nu$, Eq.~\eqref{E:gsAc2} and $\cal B$, Eq.~(\ref{E:mfcb})  one gets
\begin{eqnarray}
\langle a_m \rangle &=& \frac{t}{U} {\cal B}^{-1}\left(\frac{\mu-v_m}{U}\right) \sum_{m'} A_{m\,m'} \langle a_{m'} \rangle \nonumber\\
&=& \frac{t}{U} \sum_{m'} \Lambda_{m\,m'} \langle a_{m'} \rangle 
\end{eqnarray}
where $\Lambda$ is the same as in Eq.~\eqref{eq:Ldef}.
Recalling the criteria for the stability of linear maps, the fixed
point $\langle a_m \rangle = 0$ (equivalent to $\gamma_m = 0$) is
stable whenever
\begin{equation}
\frac{t}{U}\leq \frac{1}{|\lambda_{\max}|}
\end{equation}
where $\lambda_{\max}$ is the eigenvalue of $\Lambda$ with the largest
magnitude.

We note that despite $\Lambda^t\neq \Lambda$, the spectrum of this matrix is 
real. This can be explained by observing that the eigenvalue problem $\Lambda {\bf x} = \lambda {\bf x}$ is equivalent to $\Gamma {\bf y} = \lambda {\bf y}$, where $y_j= D_{j,j}^{-1} x_j$ and $\Gamma = \sqrt D A \sqrt D = \Gamma^t$.
Recalling that the maximal eigenvalue of a matrix coincides with its 2-norm one can derive a lower bound for the critical hopping to interaction ratio. Indeed 
\begin{eqnarray}
\frac{1}{|\lambda_{\max}|}& =& \frac{1}{\|\Lambda\|_2} \geq \frac{1}{\|D\|_2 \|A\|_2} \nonumber\\
&=& \left[2d \min_{\{v_j\}} {\cal B}^{-1}\left(\frac{\mu-v_j}{U}\right)\right]^{-1}
\label{E:magg}
\end{eqnarray}
Note that in the case of binary distributed disorder Eq.~\eqref{E:magg} coincides with Eq.~\eqref{E:finc}. Hence, the finite-size lobes always enclose their thermodynamic counterparts.

It is interesting to observe that the above approach naturally suggests two disorder-averaged effective theories. Indeed, one could trade the matrix elements of $\Lambda$ or $\Gamma$ with their averages over disorder. This would give 
\begin{equation}
\label{E:Lavg}
\frac{t}{U} < \frac{1}{2 d}  \left[\int dv\, p(v)\, {\cal B}^{-1}\left(\frac{\mu -v}{U}\right)\right]^{-1}
\end{equation}
for $\Lambda$ and 
\begin{equation}
\label{E:Gavg}
\frac{t}{U} < \frac{1}{2 d} \left[\int dv\, p(v)\, {\cal B}^{-\frac{1}{2}}\left(\frac{\mu -v}{U}\right)\right]^{-2}
\end{equation}
for $\Gamma$.

As mentioned in Ref.~\cite{Buonsante_PRA_76_011602}, Eq.~\eqref{E:Lavg} gives the zero-temperature (analytical) phase diagram derived in Refs.~\cite{Fisher_PRB_40_546,Freericks_PRB_53_2691,Krutitsky_NJP_8_187} in the case of $v$ uniformly distributed in $[-\Delta,\Delta]$. Interestingly, Eq.~\eqref{E:Gavg} gives a different result, very similar to that obtained by the  {\it stochastic mean-field theory} recently described in Ref.~\cite{Bissbort_arXiv_0804_0007}.
 
The integrations in Eqs.~\eqref{E:Lavg} and \eqref{E:Gavg} can be easily carried out analytically in the case of a binary distributed disorder, Eq.~\eqref{eq:vj}. In particular, it is quite straightforward to show that Eq.~\eqref{E:Lavg} is equivalent to Eq.~\eqref{E:smmf}, which does not exhibit the expected independence on the impuriy density $p_0=N_{\rm imp}/M$ in the thermodynamic limit, as discussed in Sec.~\ref{sec:res}. The same problem affects the boundary derived from Eq.~ \eqref{E:Gavg}. These results seem to suggest that --- at least in the case of binary disorder --- single-site mean-field theories are not able to capture the thermodynamic limit.

\section{Superfluid fraction and mean-field approach}
\label{sec:AppC}
In this section we derive the analytic expression for the phases
$\phi_i$ introduced in Section \ref{sec:MF} for the mean field study
of the currents present in the system. Also, we obtain the analytic
mean-field expression \eqref{eq:sf1mf} for the superfluid fraction \eqref{eq:sf1}.

We first of all observe that expanding the $\theta$-dependent terms in
 Eq. (\ref{eq:BHD}) one finds that even and odd contributions are purely
real and imaginary, respectively \cite{Roth_PRA_67_031602}.
As a result, this is true also of the ground-state perturbative expansion, 
while only even terms contribute to  the ground-state energy.

In the Gutzwiller approximation, this implies that each factor of the
perturbative expansion in Eq. (\ref{eq:MFgs}) has the same alternating
form: the even contributions are real and the odd contributions are
imaginary. Hence, the same property holds also for the mean-field
parameters $\alpha_m$ defined in Eq. (\ref{E:SCc}).

Therefore, at first order in $\theta$ we can write
\begin{equation}
\alpha_j = \alpha_j^0 \exp(i \theta \phi_j)
\end{equation}
where $\alpha_j^0 = \langle \psi_j^0|a_j|\psi_j^0 \rangle$ is the
local mean-field parameter for $\theta=0$, and $\phi_j \in
\mathbb{R}$.  Plugging last result in Hamiltonian (\ref{E:MFH2}) and
keeping only first-order contributions one gets
\begin{equation}
\label{e4}
{\cal H}_j= \frac{U}{2}(n_j-1)n_j + (v_j -\mu) n_j -t a_j^+ r_j \exp(i\vartheta_j)+c.c. 
\end{equation}
where we denote
\begin{equation}
\label{e7}
\vartheta_j = \theta \frac {\sum_i A_{ji} \alpha_i^0 (B_{j i}+\phi_i)}
{\sum_i A_{ji} \alpha_i^0} \quad r_j=\sum_i A_{ji} \alpha_i^0
\end{equation}
Let us define $\tilde{a}_j=a_j \exp(-i \vartheta_j)$. The new creation
and distructions operators $\tilde{a}_j^\dagger$ and $\tilde{a}_j$
satisfy the same algebra of ${a}_j^\dagger$ and ${a}_j$ and moreover
$n_j=\tilde{a}_j^\dagger \tilde{a}_j$. By introducing these new
operators the eigenvalue and the self-consistency equations become:
\begin{eqnarray}
\label{e8}
E_j |\psi_j\rangle& =& \left[ \frac{U}{2}(n_j-1)n_j + (v_j -\mu) n_j 
\right. \nonumber\\ 
& &  \left. -t (\tilde{a}_j^+ + \tilde{a}_i) \sum_i A_{j i} \alpha_j^0 
\right] |\psi_j\rangle 
\end{eqnarray}
\begin{equation}
\label{e9}
\exp(i \vartheta_j) \langle \psi_j| \tilde{a}_j|\psi_j\rangle= \alpha_j^0 \exp(i \theta \phi_j)
\end{equation}
The solutions to equations (\ref{e8}) and (\ref{e9}) can be directly
obtained from the solutions of order $0$ in $\theta$. In particular if
$|\psi_j^0\rangle= \sum_n c_{j,n} (a_j^+)^n |0\rangle $ is the on site
wavefunction for $\theta=0$, to the first order in $\theta$ the on
site wavefunctions are:
\begin{equation}
\label{e9b}
|\tilde{\psi}_j^0 \rangle = \sum_n c_{j,n} (\tilde{a}_j^+)^n |0\rangle
= \sum_n c_{j,n} e^{-i n \vartheta_j} ({a}_j^+)^n |0\rangle 
\end{equation}
with $\vartheta_j=\theta \phi_j$. Moreover, as expected, there are no
first-order contributions to the energy. Equation $\vartheta_j=\theta
\phi_j$ entails that
\begin{equation}
\label{e10}
\sum_i \left( \delta_{j i} \sum_k A_{j k} \alpha_k^0 - A_{j i}\alpha_i^0\right)\phi_i= \sum_i B_{j i} \alpha_i^0.
\end{equation}
Equations (\ref{e10}) provide an expression for the phases $\phi_i$,
which can be plugged into equation (\ref{eq:MFJ}), to obtain the
currents to the first order in $\theta$.

Let us consider the the second order perturbation in $\theta$. The
contribution to the parameters $\alpha_j$ can be written as
$\alpha_j=(\alpha_j^0+\theta^2 \xi_j)\exp(i \theta \phi_j)$. Within
such approximation the on site Hamiltonian is
\begin{equation}
\label{e11}
{\cal H}_j= 1/2(n_j-1)n_j +(v_j- \mu) n_j -t a_j^+ r'_j \exp(i \theta \phi_j)+c.c. 
\end{equation}
with
\begin{eqnarray}
r'_j  & = & \sum_i A_{j i}\alpha_i^0 + \theta^2 \left(\sum_i A_{j i}\xi_i \right.\nonumber\\
& & \left. -\frac {1}{2} \sum_i A_{j i}\alpha_i^0 (B_{j i}+\phi_i-\phi_j)^2
 \right)\label{e12}
\end{eqnarray}
Equation (\ref{e12}) has been obtained by expanding in $\theta$ and
exploiting expression (\ref{e10}). Therefore we have to solve the self
consistency equations:
\begin{equation}
\label{e13}
{\cal H}_j^0+ \theta^2 {\cal V}_j^0  |\psi_j\rangle= E_j |\psi_j\rangle
\end{equation}
and
\begin{equation}
\label{e14}
\langle \psi_j| \tilde{a}_j |\psi_j\rangle = (\alpha_j^0+\theta^2 \xi_j)
\end{equation}
where
\begin{eqnarray}
\label{e15}
{\cal H}^0_j & = & 1/2(n_j-1)n_j +(v_j-\mu) n_j \nonumber\\
& & -t (\tilde{a}_j^+ + \tilde{a}_j) \sum_i A_{ji} \alpha_j^0 \nonumber\\
{\cal V}_j & = &-t(\tilde{a}_j^+ + \tilde{a}_j) \left(\sum_i A_{ji}\xi_i \right. \nonumber\\
& & \left. -\frac {1}{2} \sum_i A_{ji}\alpha_i^0 (B_{ji}+\phi_i-\phi_j)^2\right)
\end{eqnarray}
Equations (\ref{e13}, \ref{e14}) is solved by a first order
pertubation theory in $\theta^2$. In particular denoting
$|\psi_j\rangle=|\tilde{\psi}_j^0\rangle+\theta^2
|\tilde{\psi}_j^2\rangle$ and $E_j=E_j^0+\theta^2 E_j^2$ (with
$|\tilde{\psi}_j^0\rangle$ given by (\ref{e9b}) and $E^0_i$ eigenvalue
of the solution for $\theta=0$), we have
\begin{equation}
\label{e16}
E_j^2= \langle \tilde{\psi}_j^0|  V_j  |\tilde{\psi}_j^0\rangle,
\end{equation}
\begin{equation}
\label{e18}
|\tilde{\psi}_j^2\rangle  = 
- (H_J^0- E^0_j)^{-1}  V_j  |\tilde{\psi}_j^0\rangle,
\end{equation}
and
\begin{equation}
\label{e17}
\langle \tilde{\psi}_j^0| \tilde{a}_j + \tilde{a}_j^+ |\tilde{\psi}_j^2\rangle =  \xi_j.
\end{equation}
inserting Eq. (\ref{e18}) into (\ref{e17}) we obtain
\begin{equation}
\label{e20}
\sum_i (A_{j i} X_j-\delta_{ij}) \xi_i = \frac {X_j}{2} \sum_i A_{j i}\alpha_i^0 (B_{ji}+\phi_i-\phi_j)^2
\end{equation}
with $X_j=t \langle \tilde{\psi}_j^0| (\tilde{a}_j +
\tilde{a}_j^+)(H_J^0- E^0_j)^{-1} (\tilde{a}_j + \tilde{a}_j^+)
|\tilde{\psi}_j^0\rangle$. The quantities $\xi_j$ are obtained from Eq.~(\ref{e20}) and $E_j^2$ is
\begin{equation}
\label{e21}
E_j^2= - 2 t \alpha_j^0 \left(\sum_i A_{j i}\xi_i
-\frac {1}{2} \sum_i A_{ji}\alpha_i^0 (B_{ji}+\phi_i-\phi_j)^2\right)
\end{equation}
We recall that the mean field Hamiltonian (\ref{E:MFH}) is composed
by the sum of the on site terms ${\cal H}_i$ and of a diagonal term,
which has been so far neglected, since it does not provide any
contribution to the wave functions, and hence to the relevant expectation
values. However, such term is not negligible
in the evaluation of the system energy. In particular, it provides a
contribution of $\sum_j E^d_j$ with $E^d_j={\alpha}_j^* t/2 \sum_i
A_{ji} \alpha_i \exp(i\theta B_{ji})+c.c.$. For small $\theta$ we get
$E^d_j=E^{d0}_j+\theta^2 E^{d2}_j$ with
\begin{eqnarray}
E_j^{d2} & = & t \alpha_j^0 \left(\sum_i A_{ij}\xi_i
-\frac {1}{2} \sum_i A_{ji}\alpha_i^0 (B_{ji}+\phi_i-\phi_j)^2\right)+\nonumber\\& & + t \xi_j \sum_i A_{ji}\alpha_i^0 
\label{e22}
\end{eqnarray}
from definition (\ref{eq:sf1}) it is clear that
\begin{equation}
\label{e23}
f_s=\frac{1}{t N}\sum_j E_j^2+E_j^{d2}=\frac{\sum_{i,j} A_{ij}\alpha_i^0 \alpha_j^0 (B_{ji}+\phi_i-\phi_j)^2}{2N}
\end{equation}

It is interesting to observe that on 1D systems there is no need to evaluate
the phases $\phi_j$, the superfluid fraction being determined by the $\alpha_j^0$ alone. This can be proved by observing that on a 1D lattice the current in Eq. \eqref{eq:MFJ} is bound to have the same value across any couple of neighbouring sites $j$ and $j+1$. Recalling that $A_{i j} = \delta_{i,j+1}+\delta_{i,j-1}$ and $B_{i j} = \delta_{i,j+1}-\delta_{i,j-1}$, one gets ${\cal J}_{j,j+1}=-{\cal J}_{j+1,j}={\cal J}= -2\theta t \alpha_j^0 \alpha_{j+1}^0 [\phi_j - \phi_{j+1} -1 ]$. This yields
\begin{equation}
\label{eq:J1DA}
\sum_{j=1}^M \frac{1}{\alpha^0_j\alpha^0_{j+1}}= \frac{2M  t \theta}{ {\cal J}}
\end{equation}
and
\begin{eqnarray}
\phi_j - \phi_{j+1}-1 &=& -\frac{{\cal J}}{2 \theta t \alpha_j^0 \alpha_{j+1}^0}\nonumber \\
&=& -\frac{M}{\alpha_j^0 \alpha_{j+1}^0}\left(\sum_{\ell=1}^M \frac{1}{\alpha^0_\ell\alpha^0_{\ell+1}}\right)^{-1}
\end{eqnarray}
Plugging this result into Eq. \eqref{e23} gives
\begin{equation}
\label{eq:fs1DA}
f_s = \frac{M^2}{ N} \left(\sum_{j=1}^M \frac{1}{\alpha^0_j\alpha^0_{j+1}}\right)^{-1} = \frac{\cal J}{2 t \theta N/M}.
\end{equation}

We conclude by observing that the very same derivation of Eq.~\eqref{e23} can be carried out in the case of discrete Gross-Pitaevskii equations, which, as described in Sec.~\ref{sec:App0}, ensue from assuming that the states in Eq.~\eqref{eq:MFgs} are Glauber's or SU($M$) coherent states. The resulting equations have exactly the same form as Eqs.~\eqref{e10} and \eqref{e23}, where the mean-field order parameter $\alpha_j$ is substituted by the corresponding coherent-state label $z_j$.


\end{document}